\documentclass[aps,twocolumn,noshowpacs,nobalancelastpage,tightenlines,floatfix]{revtex4}
\usepackage{amsmath}
\usepackage{amssymb}
\usepackage{graphicx}
\usepackage{dcolumn}
\usepackage{bm}

\begin{document}

\title{Single-Photon Generation from Stored Excitation in an Atomic Ensemble}
\author{C.~W. Chou, S. V. Polyakov, A. Kuzmich,$^{*}$ and H.~J.~Kimble}
\affiliation{Norman Bridge Laboratory of Physics 12-33\\
California Institute of Technology, Pasadena, CA 91125}

\begin{abstract}
Single photons are generated from an ensemble of cold Cs atoms via the
protocol of Duan et al. [Nature \textbf{414}, 413 (2001)]. Conditioned upon
an initial detection from field $1$ at $852$ nm, a photon in field $2$ at $%
894$ nm is produced in a controlled fashion from excitation stored within
the atomic ensemble. The single-quantum character of the field $2$ is
demonstrated by the violation of a Cauchy-Schwarz inequality, namely $%
w(1_{2},1_{2}|1_{1})=0.24\pm 0.05\ngeq 1$, where $w(1_{2},1_{2}|1_{1})$
describes detection of two events $(1_{2},1_{2})$ conditioned upon an
initial detection $1_{1}$, with $w\rightarrow 0$ for single photons.
\end{abstract}

\date{\today }
\pacs{PACS Numbers}
\maketitle

A critical capability for quantum computation and communication is the
controlled generation of single-photon pulses into well-defined spatial and
temporal modes of the electromagnetic field. Indeed, early work on the
realization of quantum computation utilized single-photon pulses as quantum
bits (\textit{flying qubits}), with nonlinear interactions mediated by an
appropriate atomic medium \cite{chuang95,turchette95}. More recently, a
scheme for quantum computation by way of linear optics and photoelectric
detection has been developed that again relies upon single-photon pulses as
qubits \cite{knill01}. Protocols for the implementation of quantum
cryptography \cite{lutkenhaus00} and of distributed quantum networks also
rely on this capability \cite{briegel00,duan01}, as do some models for
scalable quantum computation \cite{duan03}.

Efforts to generate single-photon wavepackets can be broadly divided into
techniques that provide photons \textquotedblleft on
demand\textquotedblright\ (e.g., quantum dots coupled to microcavities \cite%
{michler00b,moreau01,pelton02}) and those that produce photons as a result
of conditional measurement on a correlated quantum system. For conditional
generation, the detection of one photon from a correlated pair results in a
one-photon state for the second photon, as was first achieved using
\textquotedblleft twin\textquotedblright\ photons from atomic cascades \cite%
{clauser74,grangier86} and parametric down conversion \cite{hong86}, with
many modern extensions \cite{schiller01,pitman02,altepeter03,uren03}. Within
the context of the collective enhancement of atom-photon interactions in
optically thick atomic samples \cite{polzik03,lukin03}, a remarkable
protocol for scalable quantum networks \cite{duan01} suggests a new avenue
for producing single photons via conditional measurement.

\begin{figure}[tb]
\includegraphics[width=8.6cm]{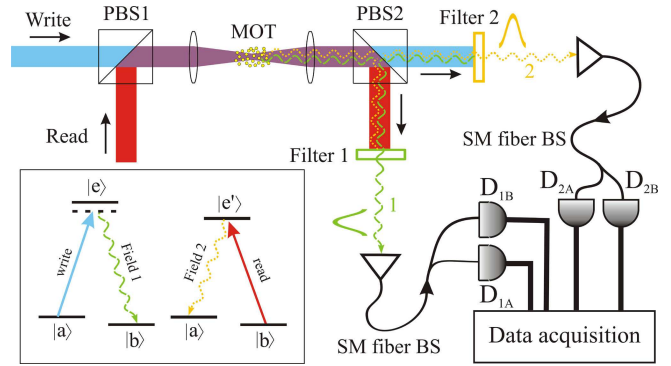}
\caption{Schematic of experiment for conditional generation of single
photons. \textit{Write} and \textit{read} pulses sequentially propagate into
a cloud of cold Cs atoms (MOT), generating the correlated output fields $%
(1,2)$. A detection event for field $1$ at $D_{1A,1B}$ leads to an
approximate one-photon state for field $2$, as confirmed with detectors $%
D_{2A,2B}$. (P)BS - (polarization) beam splitter, SM - single-mode. The
inset illustrates the relevant atomic level scheme.}
\label{setup}
\end{figure}

Inspired by the protocol of Ref. \cite{duan01}, in this Letter we report a
significant advance in the creation of single photons for diverse
applications in quantum information science, namely the generation and
storage of single quanta from an atomic ensemble. As illustrated in Figure
1, an initial \textit{write} pulse of (classical) light creates a state of
collective excitation in an ensemble of cold atoms as determined by
photoelectric detection for the generated field $1$. Although this first
step is probabilistic, its success heralds the preparation of one excitation
stored within the atomic medium. After a programmable delay $\delta t$, a
\textit{read} pulse converts the state of atomic excitation into a field
excitation, thereby generating one photon in a well-defined spatial and
temporal mode $2$. The quantum character of the $(1,2)$ fields is
demonstrated by the observed violation of a Cauchy-Schwarz inequality for
the ratio $R$ of cross correlations to auto-correlations \cite{clauser74},
namely $R=53\pm 2$ where $R\leq 1$ for any classical field \cite%
{kuzmich03,jiang03,kuzmich03-si}.

This greatly improved nonclassical correlation for photon pairs for the $%
(1,2)$ fields enables the conditional generation of single photons as in
Refs. \cite{grangier86,hong86,schiller01,pitman02,altepeter03,uren03}, but
now with the photon $2$\ stored as an excitation in the atomic ensemble \cite%
{pitman02}. Given a first photon $1$ from the \textit{write} pulse, we
trigger the emission of a second photon $2$ with the \textit{read} pulse. To
demonstrate the single-photon character of the field $2$, we measure the
three-fold correlation function $w(1_{2},1_{2}|1_{1})$ for detection of two
photons $(1_{2},1_{2})$ from field $2$ given a detection event $1_{1}$ from
field $1$, where $w(1_{2},1_{2}|1_{1})=1$ for coherent states and $%
w(1_{2},1_{2}|1_{1})\geq 1$ for any classical field. Experimentally, we find
$w(1_{2},1_{2}|1_{1})=0.34\pm 0.06$ for $\delta t=150$ ns, while $%
w(1_{2},1_{2}|1_{1})=0.24\pm 0.05$ for $\delta t=0$, thereby taking an
important step toward the creation of ideal single photons for which $%
w(1_{2},1_{2}|1_{1})\rightarrow 0$.

Figure \ref{setup} provides an overview of our experiment for producing
correlated photons from an optically thick sample of four-level atoms in a
magneto-optical trap (MOT) \cite{kuzmich03,metcalf99}. The ground states $%
\{|a\rangle ;|b\rangle \}$ correspond to the $6S_{1/2},F=\{4;3\}$ levels in
atomic Cs, while the excited states $\{|e\rangle ;|e^{\prime }\rangle \}$
denote the $\{6P_{3/2},F=4;6P_{1/2},F=4\}$ levels of the $D_{2},D_{1}$ lines
at $\{852;894\}$ nm, respectively. We start the protocol for single photon
generation by shutting off all light responsible for trapping and cooling
for $1\mu $s, with the trapping light turned off approximately $300$ ns
before the re-pumping light in order to empty the $F=3$ hyperfine level in
the Cs $6S_{1/2}$ ground state, thus preparing the atoms in $|a\rangle $.
During the \textquotedblleft dark\textquotedblright\ period, the $j^{\text{th%
}}$ trial is initiated at time $t_{j}^{(1)}$ when a rectangular
pulse of laser light\ from the \textit{write} beam, $150$ ns in
duration (FWHM) and tuned $10$ MHz below the $|a\rangle
\rightarrow |e\rangle $ transition, induces spontaneous Raman
scattering to level $|b\rangle $ via $|a\rangle \rightarrow
|e\rangle \rightarrow |b\rangle $. The \textit{write} pulse is
sufficiently weak so that the probability to scatter one Raman
photon into a forward propagating wavepacket $\psi
^{(1)}(\vec{r},t_{j}^{(1)})$ is less than unity for each pulse.
Detection of one photon\ from field $1$
results in a \textquotedblleft spin\textquotedblright\ excitation to level $%
|b\rangle $, with this excitation distributed in a symmetrized, coherent
manner throughout the sample of $N$ atoms illuminated by the \textit{write}
beam.

Given this initial detection, the stored atomic excitation can be converted
into one quantum of light at a user controlled time $%
t_{j}^{(2)}=t_{j}^{(1)}+\delta t$. To implement this conversion, a
rectangular pulse from the \textit{read }beam, $120$ ns in duration (FWHM)
and resonant with the $|b\rangle \rightarrow |e^{\prime }\rangle $
transition, illuminates the atomic sample. This pulse affects the transfer $%
|b\rangle \rightarrow |e^{\prime }\rangle \rightarrow |a\rangle $ with the
accompanying emission of a second Raman photon $2$ on the $|e^{\prime
}\rangle \rightarrow |a\rangle $ transition described by the wavepacket $%
\psi ^{(2)}(\vec{r},t_{j}^{(2)})$, where the spatial and temporal structure
of $\psi ^{(1,2)}(\vec{r},t)$ are discussed in more detail in Ref. \cite%
{duan02}. The trapping and re-pumping light for the MOT are then turned back
on to prepare the atoms for the next trial $j+1$, with the whole process
repeated at $250$ kHz.

The forward-scattered Raman light from the \textit{write}, \textit{read }%
pulses is directed to two sets of single-photon detectors ($D_{1A,1B}$ for
field $1$ and $D_{2A,2B}$ for field $2$) \cite{qe}. Light from the (\textit{%
write}, \textit{read}) pulses is strongly attenuated (by $\simeq 10^{6}$) by
the filters shown in Fig. \ref{setup}, while the associated $(1,2)$ photons
from Raman scattering are transmitted with high efficiency ($\simeq 80\%$)
\cite{kuzmich03}. Detection events from $D_{1A,1B}$ within the intervals $%
[t_{j}^{(1)},t_{j}^{(1)}+T]$ and from $D_{2A,2B}$ within $%
[t_{j}^{(2)},t_{j}^{(2)}+T]$ are time stamped (with a resolution of $2$ ns)
and stored for later analysis. $T=200$ ns for all of our measurements.

\begin{figure}[tb]
\includegraphics[width=8.6cm]{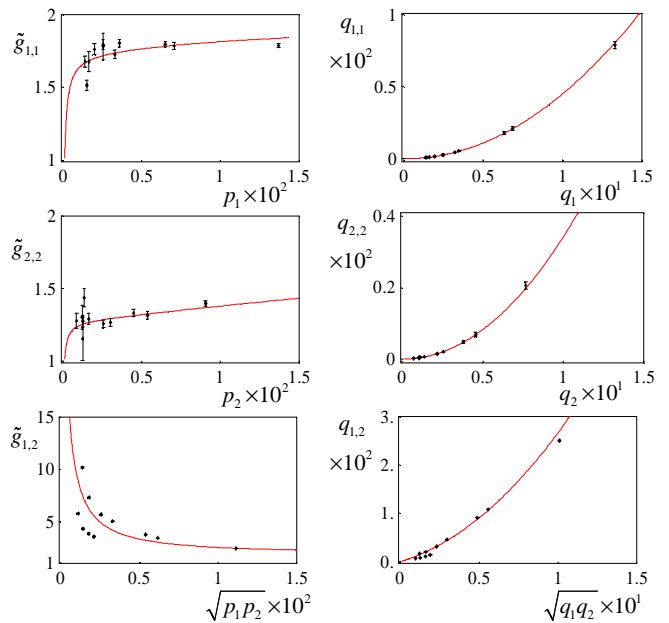}
\caption{Left column \textit{(a) -- (c)} Normalized intensity correlation
functions $\tilde{g}_{1,1},\tilde{g}_{2,2},\tilde{g}_{1,2}$ versus observed
detection probabilities $p_{1},p_{2},\protect\sqrt{p_{1}p_{2}}$,
respectively. Right column \textit{(d) -- (f)} $q_{1,1},q_{2,2},q_{1,2}$ for
joint detection versus $q_{1},q_{2},\protect\sqrt{q_{1}q_{2}}$ for single
detection, with $q_{l},q_{l,m}$ referenced to the output of the MOT.
Statistical uncertainties are indicated by the error bars. The full curves
are from the model calculation described in the text with $(\protect\kappa %
_{1},\protect\kappa _{2})=(0.17,0.90)$ and $(|v_{1b}|^{2},|v_{2b}|^{2})=0.006
$.}
\label{gijqijvspiqi}
\end{figure}

For a particular set of operating conditions, we determine the single $p_{l}$
and joint $p_{l,m}$ event probabilities from the record of detection events
at $D_{1A,1B},D_{2A,2B}$, where $(l,m)=1$ or $2$. For example, the total
singles probability $p_{1}$ for events at $D_{1A},D_{1B}$ due to field $1$\
is found from the total number of detection events $n_{1A}+n_{1B}$ recorded
by $D_{1A},D_{1B}$ during the intervals $[t_{j}^{(1)},t_{j}^{(1)}+T] $ over
$M_{tot}$\ repeated trials $\{j\}$, with then $%
p_{1}=(n_{1A}+n_{1B})/M_{tot}$. To determine $p_{1,1}$ for joint detections
at $D_{1A},D_{1B}$, we count the total number of coincidences $%
N_{1A,1B}$\ recorded by $D_{1A},D_{1B}$, with then $%
p_{1,1}=N_{1A,1B}/M_{tot} $. $p_{2,2}$ is found in an analogous
fashion using events from $D_{2A},D_{2B}$. Joint detections
between the $(1,2)$ fields are described by $p_{1,2}$, which is
determined by summing coincidence events between the four pairs of
detectors for the $(1,2) $ fields (e.g., between pairs
$D_{1A},D_{2A}$).

From $(p_{l},p_{l,m})$ we derive estimates of the normalized intensity
correlation functions $\tilde{g}_{l,m}$, where $\tilde{g}_{l,m}=1$ for
coherent states. For example, the auto-correlation function $\tilde{g}%
_{1,1}=p_{1,1}/(p_{1A}p_{1B})$ for field $1$, and similarly for the
functions $\tilde{g}_{2,2},\tilde{g}_{1,2}$ for the auto-correlation of
field $2$ and the cross correlation between fields $(1,2)$. The first column
in Figure \ref{gijqijvspiqi} displays $\tilde{g}_{1,1},\tilde{g}_{2,2}$, and $%
\tilde{g}_{1,2}$ as functions of $p_{1},p_{2},$ and $\sqrt{p_{1}p_{2}}$ \cite%
{consistency}. A virtue of $\tilde{g}_{l,m}$ is its independence from the
propagation and detection efficiencies. In the ideal case, the state for the
fields $(1,2)$ is \cite{duan01,duan02,kuzmich03-si}
\begin{equation}
|\Phi _{12}\rangle =|00\rangle +\sqrt{\chi }|11\rangle +\chi |22\rangle
+O(\chi ^{3/2})\text{ ,}  \label{phi12}
\end{equation}%
where $\sqrt{\chi }$ is the excitation amplitude for field $1$ in each trial
of the experiment. For $\chi \ll 1$, $\tilde{g}_{1,1}=\tilde{g}_{2,2}\simeq 2
$ and $\tilde{g}_{1,2}=1+1/\chi $. By contrast, for reasons that we will
shortly address, our measurements in Fig. \ref{gijqijvspiqi} give $\tilde{g}%
_{1,1}\simeq 1.7$ and $\tilde{g}_{2,2}\simeq 1.3$, with $\tilde{g}_{1,2}$
exhibiting a sharp rise with decreasing $\sqrt{p_{1}p_{2}}$, but with
considerable scatter.

To provide a characterization of the field generation that is independent of
the efficiency of our particular detection setup, we convert the
photodetection probabilities $(p_{l},p_{l,m})$ to the quantities $%
(q_{l},q_{l,m})$ for the field mode collected by our imaging system at the
output of the MOT. Explicitly, for single events for fields $(1,2)$, we
define $q_{1}\equiv p_{1}/\alpha _{1},q_{2}\equiv p_{2}/\alpha _{2}$, while
for joint events, $q_{1,1}\equiv p_{1,1}/\alpha _{1}^{2},q_{2,2}\equiv
p_{2,2}/\alpha _{2}^{2},q_{1,2}\equiv p_{1,2}/\alpha _{1}\alpha _{2}$, where
$\alpha _{l}$ gives the collection, propagation, and detection efficiency
\cite{qe}. The second column in Fig. \ref{gijqijvspiqi} displays the
measured dependence of $q_{l,m}$ for joint events versus $q_{1},q_{2},\sqrt{%
q_{1}q_{2}}$ for single events over a range of operating
conditions. As expected from Eq. \ref{phi12}, $q_{1,1},q_{2,2}$
exhibit an approximately quadratic dependence on $q_{1},q_{2}$,
while $q_{1,2}$ would be linear for $\sqrt{q_{1}q_{2}}\ll 1$ in
the ideal case.

In our experiment there are a number of imperfections that lead to
deviations from the ideal case expressed by $|\Phi _{12}\rangle $ \cite%
{duan01,kuzmich03-si,duan02}. To capture the essential aspects, we have
developed a simple model that assumes the total fields $(1,2)$ at the output
of the MOT consist of contributions from $|\Phi _{12}\rangle $, together
with background fields in coherent states $|v_{1,2}\rangle $. Operationally,
increases in $p_{1},p_{2}$ are accomplished by way of increases in the
intensity of the \textit{write} beam, with only minor adjustments to the
\textit{read} beam. Hence, we parameterize our model by taking $\chi
=|v_{w}|^{2}$, with $v_{w}$ as the (scaled) amplitude of the \textit{write}
beam. Since important sources of noise are light scattering from the \textit{%
write }and \textit{read} beams and background fluorescence from uncorrelated
atoms in the sample \cite{duan02}, we assume that $v_{1,2}=\sqrt{\kappa
_{1,2}}v_{w}$. We further allow for fixed incoherent backgrounds $%
v_{1b},v_{2b}$ to account for processes that do not depend upon increases in
the \textit{write} intensity.

\begin{figure}[tb]
\includegraphics[width=8.6cm]{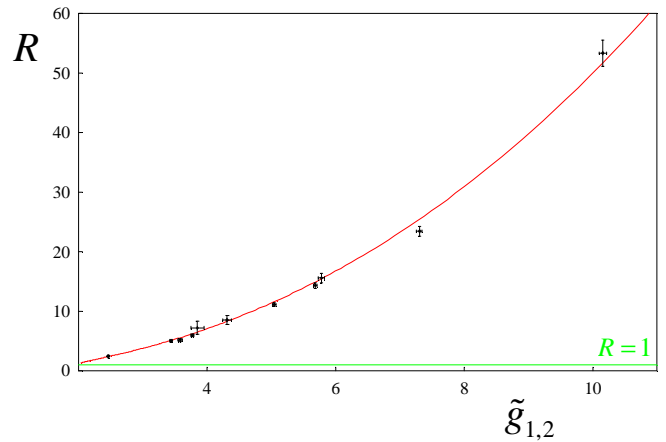}
\caption{Ratio $R\equiv \left[ \tilde{g}_{1,2}\right] ^{2}/\tilde{g}_{1,1}%
\tilde{g}_{22}$ versus the normalized cross correlation $\tilde{g}_{1,2}$,
where $R>1$ for manifestly quantum (nonclassical) fields. The points are
from our experiment with statistical uncertainties indicated by the error
bars. The full curve is from the model calculation with $(\protect\kappa %
_{1},\protect\kappa _{2})$ and $(|v_{1b}|^{2},|v_{2b}|^{2})$ as in Fig.
\protect\ref{gijqijvspiqi}.}
\label{Rvsg12}
\end{figure}

\begin{figure}[tb]
\includegraphics[width=8.6cm]{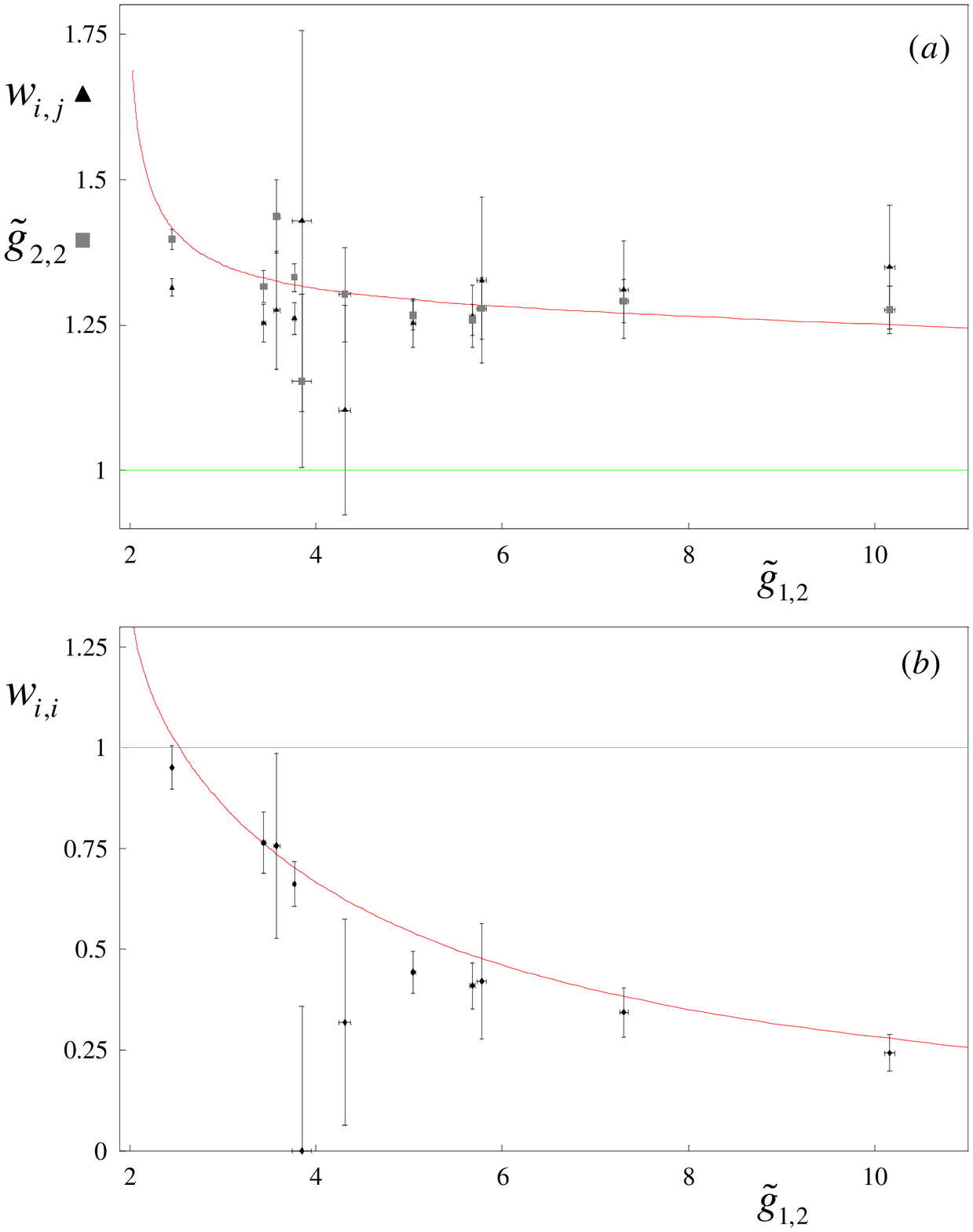}
\caption{Three-fold correlation function $w(1_{2},1_{2}|1_{1})$ for
detection event $1_{1}$ for field $1$ followed by two events $(1_{2},1_{2})$
for field $2$ versus the normalized cross correlation $\tilde{g}_{1,2}$.
\textit{(a)} $w_{i,j}(1_{2},1_{2}|1_{1})$ for events $(1_{1})_{i}$ and $%
(1_{2},1_{2})_{j}$ from different trials $i\neq j$ together with points for $%
\tilde{g}_{2,2}$. $w_{i,j}=\tilde{g}_{2,2}$ for statistically independent
trials. \textit{(b)} $w_{i,i}(1_{2},1_{2}|1_{1})$ for events from the same
trial $i$. $w_{i,i}<1$ for sub-Poissonian fields in support of the
single-photon character of field $2$. Statistical uncertainties are
indicated by the error bars. The full curves are from the model calculation
with $(\protect\kappa _{1},\protect\kappa _{2})$ and $%
(|v_{1b}|^{2},|v_{2b}|^{2})$ as in Figs. \protect\ref{gijqijvspiqi},
\protect\ref{Rvsg12}.}
\label{wvsg12}
\end{figure}

With this model, it is straightforward to compute the quantities that appear
in Figs. \ref{gijqijvspiqi}--\ref{wvsg12}. The parameters $(\kappa
_{1},\kappa _{2})=(0.17,0.90)$ and $(|v_{1b}|^{2},|v_{2b}|^{2})=0.006$ are
obtained directly by optimizing the comparison between the model results and
our measurements of normalized correlation functions (e.g., $\tilde{g}_{1,1}$
vs. $\tilde{g}_{1,2}$) without requiring absolute efficiencies. $\kappa
_{1}=0.17$ implies that the photon number for \textquotedblleft
good\textquotedblright\ events associated with $|\Phi _{12}\rangle $ exceeds
that for \textquotedblleft bad\textquotedblright\ (background) events from $%
|v_{1}\rangle $ by roughly $6-$fold for detection at $D_{1A},D_{1B}$. For
the curves in Fig. \ref{gijqijvspiqi}, we must also obtain the efficiencies $%
\beta _{l},\eta _{l}$ that convert expectation values for normally ordered
photon number operators $\hat{n}_{l}$\ for fields $l=(1,2)$ in the model
into the various $(p_{l},p_{l,m})$ and $(q_{l},q_{l,m})$ (e.g., $p_{l}=\beta
_{l}\langle \hat{n}_{l}\rangle ,q_{l}=\eta _{l}\langle \hat{n}_{l}\rangle
,q_{1,2}=\eta _{1}\eta _{2}\langle \colon \hat{n}_{1}\hat{n}_{2}\colon
\rangle $). Ideally $\beta _{l}=\alpha _{l}$ and $\eta _{l}=1$; we find
instead $(\beta _{l},\eta _{l})=(0.013,0.15)$, where we take $\beta
_{1}=\beta _{2}$ and $\eta _{1}=\eta _{2}$ for simplicity. Among various
candidates under investigation, values $\beta _{l}<\alpha _{l},\eta _{l}<1$
can arise from inherent mode mismatching for capturing collective emission
from the atomic ensemble \cite{duan02}.

Independent of the absolute efficiencies, we can utilize the results from
Fig. \ref{gijqijvspiqi} to address directly the question of the nonclassical
character of the $(1,2)$ fields by following the pioneering work of Clauser
\cite{clauser74}. The correlation functions $\tilde{g}_{l,m}$ for fields for
which the Glauber-Sudarshan phase-space function $\varphi $ is well-behaved
(i.e., \textit{classical} fields) are constrained by the inequality $R\equiv %
\left[ \tilde{g}_{1,2}\right] ^{2}/\tilde{g}_{1,1}\tilde{g}_{2,2}\leq 1$
\cite{clauser74,kuzmich03-si}. In Fig. \ref{Rvsg12} we plot the
experimentally derived values for $R$ as a function of the degree of
cross-correlation $\tilde{g}_{1,2}$ \cite{consistency}. As compared to
previous measurements for which $R=1.84\pm 0.06$ \cite{kuzmich03} and $%
R=1.34\pm 0.05$ \cite{jiang03}, we have now achieved $R\gg 1$, with $R=53\pm
2$ for the largest value of $\tilde{g}_{1,2}$. In Figs. \ref{gijqijvspiqi}
and \ref{Rvsg12} as well as \ref{wvsg12} to follow, all points are taken
with $\delta t=150$ ns, except the points at $\tilde{g}_{1,2}\simeq 10$,
which have $\delta t=0$.

This large degree of quantum correlation between the $(1,2)$ fields suggests
the possibility of producing a single photon for field $2$ by conditional
detection of field $1$. To investigate this possibility, we consider the
three-fold correlation function $w(1_{2},1_{2}|1_{1})$ for detection with
the setup shown in Fig. \ref{setup}, namely
\begin{equation}
w(1_{2},1_{2}|1_{1})\equiv \frac{p^{(c)}(1_{2},1_{2}|1_{1})}{%
[p^{(c)}(1_{2}|1_{1})]^{2}}\text{, }  \label{wdefine}
\end{equation}%
where $p^{(c)}(1_{2},1_{2}|1_{1})$ is the conditional probability for
detection of two photons $(1_{2},1_{2})$\ from field $2$ conditioned upon
the detection of an initial photon $1_{1}$ for field $1$, and $%
p^{(c)}(1_{2}|1_{1})$ is the probability for detection of one photon $1_{2}$
given a detection event $1_{1}$\ for field $1$. Bayes' theorem allows the
conditional probabilities in Eq. \ref{wdefine}\ to be written in terms of
single and joint probabilities $p^{(k)}$ for $k-$fold detection, so that%
\begin{equation}
w(1_{2},1_{2}|1_{1})=\frac{p^{(1)}(1_{1})p^{(3)}(1_{1},1_{2},1_{2})}{%
[p^{(2)}(1_{1},1_{2})]^{2}}\text{.}  \label{wjoint}
\end{equation}%
Fields with a positive-definite $\varphi $ must satisfy the Cauchy-Schwarz
inequality $w(1_{2},1_{2}|1_{1})\geq 1$. Indeed, for independent coherent
states, $w=1$, while for thermal beams, $w=2$. By contrast, for the state $%
|\Phi _{12}\rangle $ of Eq. \ref{phi12}, $w=4\chi \ll 1$ for small $\chi $,
approaching the ideal case $w\rightarrow 0$ for a \textquotedblleft
twin\textquotedblright\ Fock state $|1_{1}1_{2}\rangle $.

From the record of photo-detection events at $D_{1A,1B},D_{2A,2B}$, we
calculate estimates of the various probabilities appearing in Eq. \ref%
{wjoint}, with the results of this analysis shown in Fig. \ref{wvsg12}. Part
(a) examines the quantity $w_{i,j}(1_{2},1_{2}|1_{1})$ obtained from events
taken from different trials $i\neq j$\ for the $(1,2)$ fields (i.e.,
detection $1_{1}$ in trial $i$ for field $1$\ followed by two detections $%
(1_{2},1_{2})$ in trial $j$ for field $2$). In this case, the $(1,2)$ fields
should be statistically independent \cite{kuzmich03-si}, so that $%
w_{i,j}(1_{2},1_{2}|1_{1})=\tilde{g}_{2,2}$. Hence, we also superimpose $%
\tilde{g}_{2,2}$ from Fig. \ref{gijqijvspiqi} and find reasonable
correspondence within the statistical uncertainties (in particular, $%
w_{i,j}(1_{2},1_{2}|1_{1})\gtrsim 1$), thereby validating our analysis
techniques \cite{consistency}. No corrections for dark counts or other
backgrounds have been applied to the data in Fig. \ref{wvsg12} (nor indeed
to Figs. \ref{gijqijvspiqi}, \ref{Rvsg12}).

Fig. \ref{wvsg12} (b) displays $w_{i,i}(1_{2},1_{2}|1_{1})$ for
events from the same experimental trial $i$\ for the $(1,2)$
fields. Significantly, as the degree of cross-correlation
expressed by $\tilde{g}_{1,2}$ increases (i.e., decreasing
$\chi$), $w_{i,i}(1_{2},1_{2}|1_{1})$ drops below the classical
level of unity, indicative of the sub-Poissonian character of
the conditional state of field $2$. For $\delta t=150$ ns, $%
w_{i,i}(1_{2},1_{2}|1_{1})=0.34\pm 0.06$ for $\tilde{g}_{1,2}=7.3$, while
with $\delta t=0$, $w_{i,i}(1_{2},1_{2}|1_{1})=0.24\pm 0.05$ for $\tilde{g}%
_{1,2}=10.2$. Beyond the comparison to our model shown the figure,
empirically we find that $w_{i,i}(1_{2},1_{2}|1_{1})$ is well approximated
by $\tilde{g}_{1,1}\tilde{g}_{2,2}/\tilde{g}_{1,2}$, as in the ideal case of
Eq. \ref{phi12}. However, independent of such comparisons, we stress that
the observations reported in Fig. \ref{wvsg12} represent a sizable
nonclassical effect in support of the conditional generation of single
photons for field $2$.

In conclusion, our experiment represents an important step in the creation
of an efficient source of single photons stored within an atomic ensemble,
and thereby towards enabling diverse protocols in quantum information
science \cite{knill01,lutkenhaus00,duan01,duan03}. Our model supports the
hypothesis that the inherent limiting behavior of $w_{i,i}(1_{2},1_{2}|1_{1})
$ below unity is set by the efficiency $\eta _{l}$, which leads to
prohibitively long times for data acquisition for $\chi \lesssim 0.04$,
corresponding to the smallest value of $w_{i,i}$ in Fig. \ref{wvsg12}. We
are pursuing improvements to push $\eta _{l}\simeq 0.15\rightarrow 1$,
including in the intrinsic collection efficiency following the analysis of
Ref. \cite{duan02}. Dephasing due to Larmor precession in the quadrupole
field of the MOT limits $\delta t\lesssim 300$ ns, which can be extended to
several seconds in optical dipole or magnetic traps \cite{metcalf99}.

We gratefully acknowledge the contributions of A. Boca, D. Boozer, W. Bowen,
and L.-M. Duan. This work is supported by ARDA, and by the Caltech MURI
Center for Quantum Networks and by the NSF.

$^{*}$ School of Physics, Georgia Institute of Technology, Atlanta, Georgia
30332

\newpage


\begin{thebibliography}{99}
\bibitem{chuang95} I. L. Chuang and Y. Yamamoto, Phys. Rev. A 52, 3489
(1995).

\bibitem{turchette95} Q. A. Turchette \textit{et al.}, \textit{Phys. Rev.
Lett.} \textbf{75}, 4710-4713 (1995).

\bibitem{knill01} E. Knill, R. Laflamme, and G. Milburn, \textit{Nature}
\textbf{409}, 46-52 (2001).

\bibitem{lutkenhaus00} N. Lutkenhaus, Phys. Rev. A \textbf{61}, 052304
(2000).

\bibitem{briegel00} H.-J. Briegel and S. J. van Enk, in \textit{The Physics
of Quantum Information}, eds. D. Bouwmeester, A. Ekert, and A.
Zeilinger (Springer-Verlag, Berlin, 2000), \textbf{6.2} \&
\textbf{8.6}.

\bibitem{duan01} L.-M. Duan, \textit{et al.}, \textit{Nature} \textbf{414},
413 (2001).

\bibitem{duan03} L.-M. Duan and H. J. Kimble, quant-ph/0309187.

\bibitem{michler00b} P. Michler \textit{et al.}, Science \textbf{290}, 2282
(2000).

\bibitem{moreau01} E. Moreau \textit{et al.}, Appl. Phys. Lett. \textbf{79},
2865 (2001).

\bibitem{pelton02} M. Pelton \textit{et al.}, Phys. Rev. Lett. \textbf{89},
233602 (2002).

\bibitem{clauser74} J. F. Clauser, Phys. Rev. D \textbf{9}, 853 (1974).

\bibitem{grangier86} P. Grangier, G. Roger, and A. Aspect, Europhys. Lett.
\textbf{1}, 173 (1986).

\bibitem{hong86} C. K. Hong and L. Mandel, \textit{Phys. Rev. Lett.} \textbf{%
56}, 58 (1986).

\bibitem{schiller01} A. I. Lvovsky \textit{et al.}, \textit{Phys. Rev. Lett.}
\textbf{87}, 050402 (2001).

\bibitem{pitman02} T. B. Pittman, B. C. Jacobs, J. D. Franson, \textit{Phys.
Rev. A}\textbf{66}, 042303 (2002).

\bibitem{altepeter03} J. B. Altepeter \textit{et al.}, \textit{Phys. Rev.
Lett.} \textbf{90}, 193601 (2003).

\bibitem{uren03} A. B. U'Ren \textit{et al.}, quant-ph/0312118.

\bibitem{polzik03} B. Julsgaard \textit{et al.}, \textit{Q. Inf. \&
Computation} \textbf{3}, 518 (2003).

\bibitem{lukin03} M. Lukin, \textit{Rev. Mod. Phys. }\textbf{75}, 457 (2003).

\bibitem{kuzmich03} A. Kuzmich \textit{et al}., Nature \textbf{423}, 731
(2003).

\bibitem{jiang03} Wei Jiang \textit{et al.}, quant-ph/0309175.

\bibitem{kuzmich03-si} See Supplementary Information accompanying Ref. \cite%
{kuzmich03}.

\bibitem{metcalf99} \textit{Laser Cooling and Trapping}, H. J. Metcalf and
P. van der Straten (Springer-Verlag, 1999).

\bibitem{duan02} L.-M. Duan, J. I. Cirac, and P. Zoller, \textit{Phys. Rev.
A }\textbf{66}, 023818 (2002).

\bibitem{consistency} As a consistency check, we have employed white light
for measurements as in Figs. $2-4$, and find that $\tilde{g}_{1,1}=1.02\pm
0.01,\tilde{g}_{2,2}=1.01\pm 0.01,\tilde{g}_{1,2}=1.02\pm
0.01,w_{i,i}=0.99\pm 0.2,w_{i,j}=0.97\pm 0.02$, where in all cases, these
correlation functions should equal unity.

\bibitem{qe} The overall efficiencies $\alpha _{1,2}=\xi
_{1,2}T_{1,2}\varsigma _{1,2}$, where $\xi _{1,2}=(0.41\pm 0.04,0.47\pm
0.04) $ for light with the spatial shape of the \textit{write}, \textit{read}
beams propagating from the MOT to the input beam splitters for detectors $%
(D_{1A,1B},D_{2A,2B})$, which have quantum efficiencies $\varsigma
_{1,2}\simeq (0.50,0.40)$ (i.e., photon \textit{in} to TTL pulse \textit{out}%
). The efficiencies $T_{1,2}=0.50$ for PBS2 in Fig. \ref{setup}
account for the presumed unpolarized character of the $(1,2)$
fields in our experiment.
\end{thebibliography}
\end{document}